\documentstyle[prl,aps,epsf]{revtex}
\tightenlines
\begin{document}
\voffset=1cm \draft
\title{Giant-dipole Resonance and the Deformation
of Hot, Rotating Nuclei}
\author{Dimitri Kusnezov$^a$ and W. Erich Ormand$^b$}
\address{$^a$Center for Theoretical Physics, Sloane Physics Laboratory,\\
Yale University, New Haven, CT 06520-8120\\
$^b$ Lawrence Livermore National Laboratory, L-414, P.O. Box 808,
Livermore, CA 94551} \maketitle
\begin{abstract}
The development of nuclear shapes under the extreme conditions of
high spin and/or temperature is examined. Scaling properties are 
used to demonstrate
universal properties of both thermal expectation values of nuclear
shapes as well as the minima of the free energy, which can be used
to understand the Jacobi transition.  A universal
correlation between the width of the giant dipole resonance and
quadrupole deformation is found, providing a novel probe to
measure the nuclear deformation in hot nuclei.

\end{abstract}
The study of the properties of the giant-dipole resonance
(GDR) at finite excitation energy (or temperature, $T$) and/or
angular momentum,$J$, provides an insight into the behavior
of nuclei under extreme conditions~\cite{r:1}. A wealth of
experimental data now exist for nuclei ranging from
44~$\le$~A~$\le$~208 (see for example
Refs.~\cite{r:1,r:2,r:3,r:4}) detailing the properties of the
GDR centroid $E_0$ and damping width $\Gamma$ with
excitation energy and angular momentum of the compound nucleus.
These experiments provide an important testing ground for
theoretical models of the GDR; in particular, the role played by
quantal and thermal fluctuations in the damping of the giant
vibration. Based on both qualitative and quantitative
agreement with experimental data, the thermal fluctuation model
(TFM)~\cite{r:5,r:6,r:7,r:7a} provides a coherent picture of the GDR in
nuclei at finite excitation energy and/or angular momentum. A
systematic study of the thermal fluctuation model has recently
revealed the existence of a universal scaling law for the width of
the GDR for all $T$, $J$, and $A$~\cite{r:8}.

An important feature of the thermal fluctuation model is that the
GDR strength function is directly related to the sampling of the
quadrupole deformation in the hot nucleus. This permits an 
examination of the relationship between the damping of the GDR and
the mean quadrupole deformation. Our aim is to understand
whether simple, universal behaviors exist for the evolution of
nuclear shapes and their fluctuations. We also investigate the
systematics of the free energy surfaces through the location of
the minima $\beta_0$, $\gamma_0$,
the deformation parameters $\beta$ and $\gamma$, including their
mean values $\langle\beta\rangle$ and $\langle\gamma\rangle$ and
their variances $\langle\beta^2\rangle$ and
$\langle\gamma^2\rangle$, as a function of $T$, $J$, $A$, and
$Z$. The universality we find for $\beta$ and $\gamma$
allows the tracing out of the Jacobi transition at high
angular momentum. We further find a universal relationship between
the full-width-at-half-maximum (FWHM) of the GDR strength function 
and the mean quadrupole deformation $\langle\beta\rangle$. Consequently, a
single measurement of the FWHM can be simply related to the mean deformation 
of the system and readily extracted from experiment.

The thermal fluctuation model is based on the fact that
large-amplitude thermal fluctuations of the nuclear shape play an
important role in describing observed nuclear properties. If
the time scale associated with thermal fluctuations
is slow compared to the shift in the dipole frequency caused by
the fluctuations (adiabatic motion), the observed GDR strength
function can be obtained from a weighted average over all nuclear
shapes and orientations. Projecting the z-component, $J_z$, of
angular momentum, the GDR cross section is given by~\cite{r:9}
\begin{equation}
\sigma(E) = Z_{J}^{-1} \int
\frac{{\cal D}[\alpha]}{{\cal I}(\beta,\gamma,\theta,\psi)^{1/2}}
 \sigma(\vec\alpha,\omega_{J};E)
{\rm e}^{-F(T,\vec\alpha,J_z)/T},
\label{e:avg_j}
\end{equation}
where ${\cal D}[\alpha]=
\beta^4d\beta\sin(3\gamma)d\gamma\sin\theta d\theta d\phi d\psi$, 
$E$ is the photon energy, $Z_J=\int {\cal
D}[\alpha]/{\cal I}^{1/2}{\rm e}^{-F/T}$, 
${\cal I}$ is a function dependent on the 
principal moments of inertia and the Euler angles, and $F$ is the free
energy at finite angular momentum. 
As with previous applications of the adiabatic picture, 
we model the GDR as a rotating, three-dimensional harmonic oscillator. 
Details describing the evaluation of $\sigma(\vec\alpha,\omega_{J};E)$
are given in Ref.~\cite{r:7a}. 

The free energy is evaluated using the
Nilsson-Strutinsky~\cite{r:Str66} procedure and its extensions to
finite temperature~\cite{r:Bra81}. It is composed of two
components,
\begin{equation}
F=F_{LD}+F_{SH},
\end{equation}
where $F_{LD}$ is the free energy of a liquid-drop and $F_{SH}$ is the
Strutinsky shell correction. Shell corrections play an important
role for well-deformed and doubly-magic nuclei at low 
temperature and exhibit no known
global or universal behaviors. On the other hand, shell corrections weaken
with increasing temperature, and most non-doubly magic nuclei
have essentially {\it melted} by $T = 1.25$~MeV. For doubly magic nuclei, such
as {$^{208}$Pb, the shell corrections at $T=1.5$~MeV have diminished by a
factor of ten from their value at $T=0$~\cite{r:7,r:7a}. Consequently, any
derived universal scaling law using liquid-drop free energies should be
applicable to non-doubly magic nuclei for $T \ge 1.25$~MeV and essentially all
nuclei for $T \ge 1.75$~MeV. In this work, we utilize a liquid-drop
parameterization based on finite-temperature, extended Thomas-Fermi
calculations~\cite{r:Gue88}. This liquid-drop parameterization includes the
curvature term and all coefficients are temperature-dependent

In this work, we focus on the mass region $A=44-208$. For the 
GDR parameters 
defining the central frequency for spherical shapes, $E_0$, and the intrinsic
width $\Gamma_0$ defined in Ref.~\cite{r:7a} we use: 
$(E_0=22,\Gamma_0=4)$ for $^{44}$Ti, $(17,4)$ for $^{90}$Zr,
$(15.5,5)$ for $^{120}$Sn, $(14,4)$ for $^{168}$Er and $(13.6,4)$
for $^{208}$Pb, all in MeV.

We now examine in more detail the
behavior of the nuclear shape under extreme conditions. 
Using Eq.~(\ref{e:avg_j}), we define the mean
value of any observable, and in particular moments of the
quadrupole deformation parameter $\beta$ (and similarly for $\gamma$) as
\begin{equation}\label{bav}
\langle\beta^n\rangle = Z_{J}^{-1} \int \frac{{\cal D}[\alpha]}{{\cal
I}(\beta,\gamma,\theta,\psi)^{1/2}}
 \beta^n {\rm e}^{-F(T,\vec\alpha,J_z)/T}.
\end{equation}
The spin dependence of the moments
can be isolated by studying the normalized
expectation values,
\begin{equation}\label{vevs}
\frac{\langle\beta^n(J,T,A)\rangle}{\langle\beta^n(J=0,T,A)\rangle},\qquad
\frac{\langle\gamma^n(J,T,A)\rangle}{\langle\gamma^n(J=0,T,A)\rangle}
\end{equation}
at a fixed temperature $T$. These are shown in Figs.~\ref{fig:fig1}(a-d) 
as a function of $\xi$ for
$T=1$ MeV and selected nuclei, displaying scaling. 
When we repeat this for other values of
$T$, we obtain a family of distinct curves. (We
will focus now on $\langle\beta\rangle$ and $\langle\beta^2\rangle$ to
simplify the discussion.) Taking the same $T$-dependent power law
as in \cite{r:8}, we can remove the $T$ dependence of the curves
in Figs.~\ref{fig:fig1}(a-d) by considering
$[\langle\beta^n(J,T,A)\rangle/\langle\beta^n(J=0,T,A)\rangle]^{(T+3)/4}$,
which now becomes both mass and temperature independent as seen in
Figs.~\ref{fig:fig2}(a-b) (symbols). The results are the universal functions
(solid) for each $n$ and observable. We denote these $w_n(\xi)$
for $\beta$; shown as the solid lines in Figs.~\ref{fig:fig2}(a-b). The final
quantity for these thermal fluctuation results is the $J=0$
contribution to the average deformation. $\langle\beta^n(J,T,A)\rangle$
at $J=0$ is estimated from Eq.~(\ref{bav}) to be approximately
\begin{equation}
 \langle\beta^n(J=0,T,A)\rangle= \frac{4}{3\sqrt\pi}\; \Gamma
 \left(\frac{n+5}{2}\right)\left(\frac{T}{C_{0}(Z,A)}\right)^{n/2},
\end{equation}
where 
$C_{0}(Z,A) \simeq (1/2)\partial^2F/\partial\beta^2\mid_{\beta=0}.$
The $Z$ and $A$ dependence of $C_{0}$ can be estimated
from the dominant contributions to the liquid drop
energy\cite{r:Hass}, giving 
\begin{equation}\label{ceff1}
  C_{\scriptsize 0}(Z,A)=a_1 A^{2/3} + a_2 \frac{(N-Z)^2}{A} + 
a_3\frac{Z^2}{A} +  a_4  \frac{Z^2}{A^{1/3}}.
\end{equation}
The calculated $\langle\beta^n(J=0,T,A)\rangle$ values are well reproduced
with $a_1=3.09$, $a_2=-0.74$, $a_3=0.12$, $a_4=-0.066$, as seen
in Fig.~\ref{fig:fig3}.
These values compare quite well to the coefficients
of the liquid-drop Free energy. Here we point out an interesting
distinction between the $J=0$ temperature dependence of
$\langle\beta\rangle$ and the GDR width $\Gamma$: the former has
$T^{1/2}$ behavior while the latter has been found to behave as
$\log(1+T)$~\cite{r:8}.

Gathering the results, we have
\begin{eqnarray}\label{bfin}
  \langle\beta(J,T,A)\rangle &=& \frac{8}{3\sqrt\pi}\sqrt{\frac{T}{C_{0}}}
  w_1(\xi)^{4/(T+3)},\\
\label{e:eq8}
  \langle\beta^2(J,T,A)\rangle &=& \frac{5T}{2C_{0}}
  w_2(\xi)^{4/(T+3)}.
\end{eqnarray}
While not unique, we follow \cite{r:8} and parameterize $w_n$ as
shifted Fermi-functions

\begin{eqnarray}\label{fermi}
 w_1(\xi)=1+\frac{a_1}{1+\exp[(a_2-\xi)/a_3]},\quad w_2(\xi)=w_1(\xi)^2
\nonumber
\end{eqnarray}
where $a_i=(4.3,1.64,0.31)$. Analogous functions can be found for 
$\langle\gamma^n\rangle$ as is evident from Fig.~\ref{fig:fig1}. 
Eq.~(\ref{bfin}) describes the mean nuclear
deformation of a hot nucleus of mass $A$, temperature $T$ and spin
$J$, demonstrating that the nucleus not only undergoes a smooth
evolution under extreme conditions, but that the results are {\it
generic}, with a simple function describing all the systematics.

While the same analysis is readily done for other observables, let us
consider properties of the free energy surface itself, such as
the location of the minimum of $F$ as a function of
$J,T,A$, denoted by $\beta_{min},\gamma_{min}$. The location of
the minimum is shown in Fig.~\ref{fig:fig2}(right) for selected nuclei, at
$T=1-4$ MeV. We find that when plotted versus $\xi$, these display
the same universality. The Jacobi transition now becomes evident 
at $\xi\sim 1.2$,
where $\gamma_{min}$ displays an abrupt change from $\pi/3$ to 0. We can 
use this to define a critical spin for the transition: $J_c\sim 1.2 A^{5/6}$. 
We note that  although the curve plotted is for $T=1$~MeV, little variation 
is exhibited for other values of $T$. In Fig.~\ref{fig:fig4}, 
we plot the generic 
evolution of the free energy surface minimum as a function of $\xi$ 
that arises in the TFM.

We have seen that simple scaling
properties emerge for the average deformations and their powers.
Taken together with the scaling previously seen in the GDR width
$\Gamma$, it follows that there should be some functional relation
between the average deformation $\langle\beta\rangle$ and $\Gamma$. This
is likely since the primary mechanism for spreading the GDR modes is
a vibrational dephasing caused by the shift in the centroids of the 
GDR modes at each deformation, which to first order is a linear 
function in $\beta$. Hence, we propose the ansatz 
\begin{equation}\label{e:wid}
  \langle\beta(J,T,A)\rangle = 
a^\prime\frac{\Gamma(J,T,A)-\Gamma_0}{E_0}+c^\prime,
\label{e:beta_FWHM}
\end{equation}
and find remarkable agreement 
for all nuclei at all temperatures and 
spins below the Jacobi transition with $a^\prime=0.8$ and 
$c^\prime=0.12$ as illustrated in Fig.~\ref{fig:fig5}.
Indeed, we see that for all nuclei, 
the average deformation depends {\it linearly} on the
FWHM  and is reproduced
by Eq.~(\ref{e:beta_FWHM}) with an average fluctuation, or uncertainty,
of $\delta\langle\beta\rangle\approx \pm 0.05$. Hence, with the 
FWHM one can probe the evolution of the nuclear 
shape directly from experiment.

Eq.~(\ref{e:wid}) provides a convenient and direct experimental measure of 
nuclear deformation. In Fig.~\ref{fig:fig6}
recent experimental data on Sn~\cite{r:1,r:2,r:3,r:4} is used to extract
empirical deformations. The error bars in the figure were obtained by adding
in quadrature the quoted 
experimental errors~\cite{r:1,r:2,r:3,r:4} and the 
uncertainty $\delta\langle\beta\rangle$ 
from Eq.~(\ref{e:beta_FWHM}).
The values $\langle\beta\rangle$ deduced from experimental data are also 
compared directly to TFM calculations (solid) as well as scaling law 
predictions (dashes) of Eq.~(\ref{bfin}). 
One can see that Eq.~(\ref{e:wid}) provides a 
convenient measure of both the $J$ and $T$ 
dependence of the deformation, and is readily computed from
the data.

It has recently been argued that the spin-dependence of the GDR 
width $\Gamma$ in $^{194}$Hg is not seen experimentally since the
fluctuations $\sigma_\beta = \sqrt{\langle\beta^2\rangle -\langle\beta
\rangle^2}$ are larger in magnitude than $\beta_{min}$, so that
rotational effects due to the deformation are washed out\cite{r:cam}.
Using (7)-(8), Fig. 2, and $C_0(A,Z)$ for $^{194}$Hg, we compute 
that $\sigma_\beta\sim \beta_{min}$ when $J\sim 37\sqrt{T}$. At 
$T=1.3$~MeV, $J<42$ agrees with observations and predictions in 
\cite{r:cam}. This condition complements the analysis of \cite{r:8},
providing a deformation-based, analytic approach to understanding the
observed experimental features of the GDR.

In conclusion, within the framework of the thermal fluctuation
model, we have studied new universal scaling properties. 
We find a simple scaling behavior for all nuclei at all 
temperatures and spins for the moments of parameters defining the 
quadrupole deformation, $\beta$ and $\gamma$. In addition, we find a 
remarkable linear relationship between the 
FWHM of the GDR and mean value of the quadrupole deformation, 
$\langle\beta\rangle$. Consequently, the FWHM may be viewed as an experimental
probe of the the evolution of the nuclear shape.

We thank K. Snover for useful discussions.
This work was performed in part under auspices of 
the U. S. Department of Energy by the University of California,
Lawrence Livermore National Laboratory under contract 
No. W-7405-Eng-48.
\bibliographystyle{try}

\begin{figure}
\leavevmode \epsfxsize=8cm\epsfbox{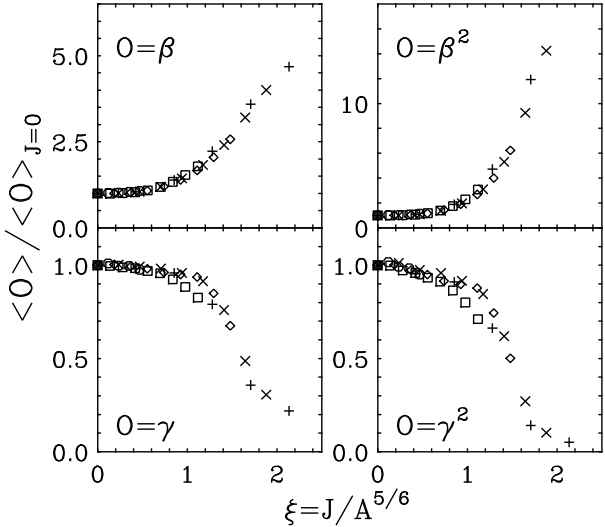}
\caption{Dependence of
normalized shape parameters as a
function of $\xi=J/A^{5/6}$ for $^{44}$Ti $(\times)$, $^{90}$Zr
$(+)$, $^{120}$Sn $(\Diamond)$, $^{168}$Er $(\Box)$, $^{208}$Pb
(o) at $T=1$~MeV, demonstrating scaling for fixed $T$.}
 \label{fig:fig1}
\end{figure}

\begin{figure}
\leavevmode \epsfxsize=8cm\epsfbox{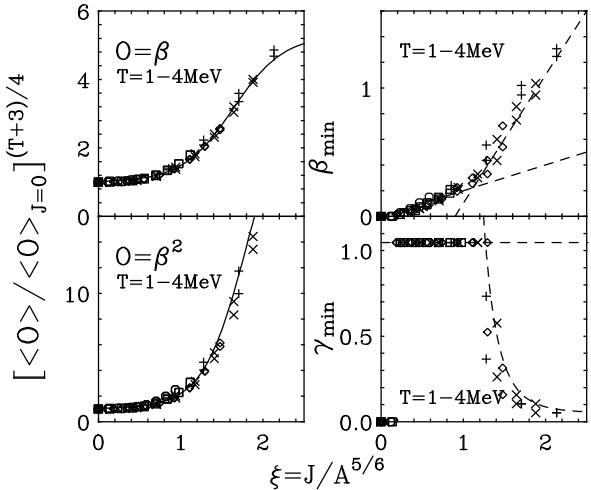}
\caption{$\xi$ dependence of: (Left) normalized expectation values of 
deformation
from $T=1-4$Mev for $^{44}$Ti $(\times)$, $^{90}$Zr $(+)$,
$^{120}$Sn $(\Diamond)$, $^{168}$Er $(\Box)$, $^{208}$Pb (o)
indicating $T$ independence. Solid lines are Eqs. (4)-(9);
(Right) Location of free energy minima
$(\beta_{min},\gamma_{min})$ for same nuclei as a function of
$\xi$ for $T=1-4$ MeV. Dashed lines are for guidance. The Jacobi transition
is evident at $\xi\sim 1.2$. Below that, $\beta_{min}\sim \xi/5$ and 
$\gamma_{min}\sim 1$.}
\label{fig:fig2}
\end{figure}

\begin{figure}
\leavevmode \epsfxsize=8cm\epsfbox{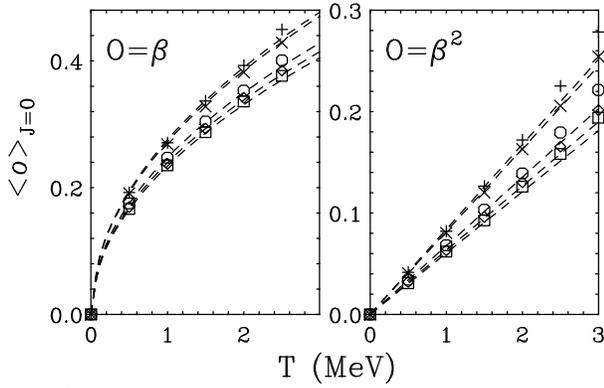}
\caption{Temperature
dependence of moments of deformation at $J=0$ (symbols) compared
to predictions from Eqs. (5)-(6). Symbols are: $^{44}$Ti $(\times)$, 
$^{90}$Zr $(+)$,$^{120}$Sn $(\Diamond)$, $^{168}$Er $(\Box)$, 
$^{208}$Pb (o). Note that at a given $T$, the ordering of the symbols is
not monotonic with $A$, and is well reproduced by Eq.~(5).}
\label{fig:fig3}
\end{figure}

\begin{figure} 
\leavevmode \epsfxsize=8cm\epsfbox{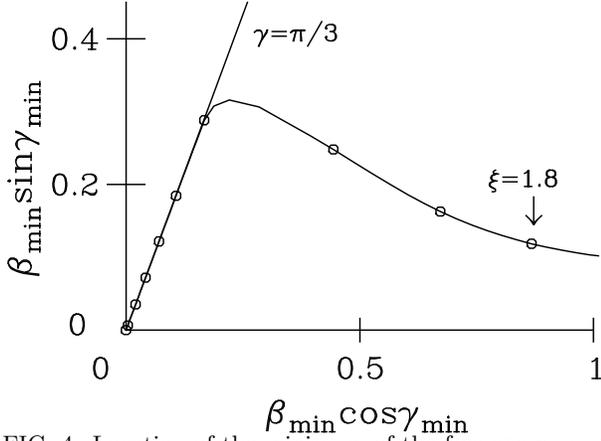}
\caption{Location of the minimum of the free energy surface for
hot, rotating nuclei. The circles are shown in increments of
$\xi=0.2$, starting with $\xi=0$ at the origin.}
\label{fig:fig4}
\end{figure}

\begin{figure}
\leavevmode \epsfxsize=8cm\epsfbox{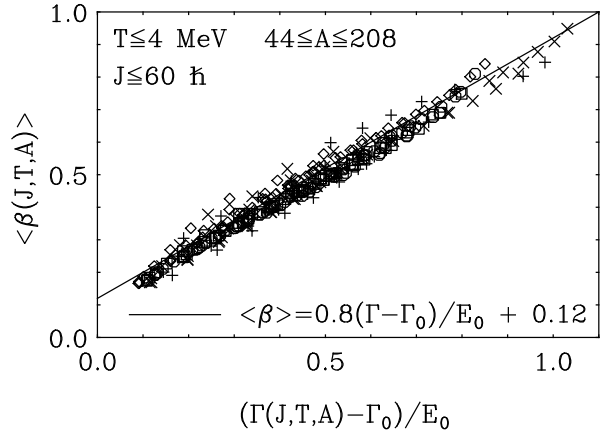}
\caption{Linear relation of average nuclear deformation to the GDR
width $\Gamma$, for $44\le A\le 208$, $T\le 4$~MeV and $J\le
60\hbar$, illustrating that $\Gamma$ can be used as a direct
experimental measure of nuclear deformation for any $J$, $T$, $A$
where shell corrections are unimportant. Symbols represent the same
nuclei as in Figs.~\ref{fig:fig1}-\ref{fig:fig3}.}
\label{fig:fig5}
\end{figure}
\begin{figure}[h!]
\epsfxsize=8cm\epsffile{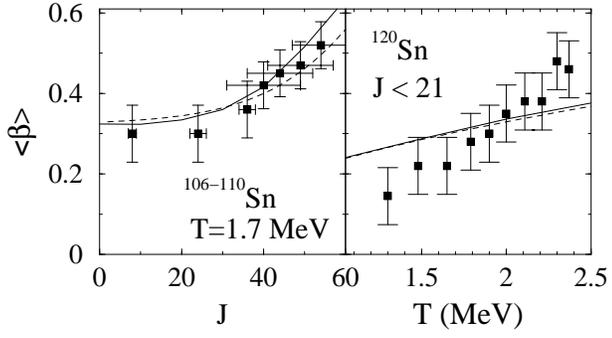}
\caption{Average nuclear deformation $\langle\beta\rangle$ extracted from
experiment~[1-4]
using Eq.~(\ref{e:beta_FWHM}), as a 
function of $J$ and $T$. Comparison 
to exact TFM calculations (solid) and scaling predictions of 
Eq.~(\ref{bfin}) (dashes) 
are quite good.}
\label{fig:fig6}
\end{figure}

\end{document}